\documentclass[final,5p,times,twocolumn]{elsarticle}

\usepackage{amsmath,amssymb,bm}
\usepackage{graphicx}
\usepackage{xcolor}
\usepackage{booktabs}
\usepackage{hyperref}
\usepackage{siunitx}

\journal{Physics Letters B}

\newcommand{\dd}{\mathrm{d}}

\newcommand{\mc}{\mathcal}

\newcommand{\Efour}{E/(4\pi)}

\makeatletter
\let\ps@pprintTitle\ps@plain
\makeatother

\begin{document}
\begin{frontmatter}

\title{
Metastable and critical-bubble branches of Coleman--Weinberg monopoles}

\author[a,b]{Sumit Shaw}
\ead{sumitshaw@imsc.res.in}
\address[a]{
The Institute of Mathematical Sciences, Chennai 600113, India
}
\address[b]{Homi Bhabha National Institute, Training School Complex, Anushaktinagar, Mumbai 400094, India}

\begin{abstract}
We revisit the Coleman--Weinberg monopole problem introduced by Kiselev,
where radiative symmetry breaking makes the broken vacuum metastable.
We construct the associated static monopole--critical-bubble
configuration in the full coupled radial Higgs--gauge system and show
that it is a saddle of the static energy functional.  The metastable
monopole and monopole--critical-bubble branches are characterized by
their profiles, energies, and radial Hessian spectra.  The
monopole--bubble solution carries a negative radial mode, while the metastable monopole remains 
locally stable until its lowest radial Hessian eigenvalue approaches zero.
The resulting branch structure gives a direct static picture of
how Coleman--Weinberg monopoles lose metastability, with
critical rescaled scalar mass parameter \(\mu_c=0.064352(1)\).
\end{abstract}


\end{frontmatter}

\section{Introduction}
\label{sec:intro}

Magnetic monopoles in Yang--Mills--Higgs theory are canonical
finite-energy solitons in relativistic field theory.  
The regular monopole of 't Hooft and Polyakov~\cite{tHooft1974,Polyakov1974} is
supported by the topology of the symmetry-broken vacuum, while the
Bogomolny--Prasad--Sommerfield limit provides analytic control over its
structure~\cite{PrasadSommerfield1975,Bogomolny1976}. 
More broadly, monopoles have become a standard setting for nonperturbative field theory, topology, and semiclassical physics~\cite{GoddardOlive1978,
Preskill1984,Rajaraman1982,MantonSutcliffe2004,VilenkinShellard1994,
Bais2004}.

A different question arises when the symmetry-broken phase is only metastable. 
In that case the monopole is not merely a topological object
embedded in a fixed vacuum: its core can probe the lower-energy region of field space and seed vacuum decay.  
This connects monopole physics with false-vacuum decay and critical bubbles~\cite{KobzarevOkunVoloshin1975,
Coleman1977,CallanColeman1977,Langer1967,Affleck1981,Linde1981,
Linde1983}, as well as with defect-assisted transitions~\cite{
Steinhardt1981,PreskillVilenkin1993,KumarParanjapeYajnik2010,
AgrawalNee2022,ParanjapeSaxena2024}.  
The relevant static object is a critical configuration on the barrier between a localized metastable monopole and an expanding bubble.  
Such a configuration is a saddle of the energy functional and cannot be obtained by ordinary dissipative
relaxation.

The Coleman--Weinberg monopole problem was first studied by
Kiselev~\cite{Kiselev1990}.  
There the Higgs expectation value is
generated radiatively, and the broken vacuum becomes metastable for sufficiently small scalar mass.  
Kiselev argued that the ordinary monopole branch is accompanied by another stationary configuration. 
This is interpreted as a monopole superposed with a critical bubble, using a reduced large-distance equation and a branch diagnostic denoted by \(C_2\).  
To our knowledge, however, the full radial \((H,K)\)
monopole--critical-bubble profile has not been explicitly
constructed as a coupled boundary-value solution with a direct Hessian negative-mode check.  
Related recent work has also shown that Coleman--Weinberg
corrections can qualitatively modify soliton structure and scales~\cite{
EtoHamadaJinnoNittaYamada2022,Kim2024}.

This work provides such a construction, by solving the full coupled radial equations for the Higgs profile \(H(r)\) and gauge profile \(K(r)\) in the
Coleman--Weinberg potential.
We obtain the monopole--critical-bubble saddle,
and characterize it through its energy, radius, and radial Hessian spectrum.  
We also follow the ordinary monopole branch to its endpoint,
where the lowest radial Hessian mode approaches zero.  
The result is the realization of the branch structure anticipated in ~\cite{Kiselev1990}, together with a clarification of the normalization entering the reduced large-distance comparison.

The present work is restricted to static, spherically symmetric
configurations.  
We construct the radial critical solution and determine its Morse character, but do not compute the Euclidean time-dependent bounce, the real-time decay, or non-radial fluctuations.  
The results may provide the foundation for the monopole-catalysed tunnelling problems considered in ~\cite{
Steinhardt1981,PreskillVilenkin1993,KumarParanjapeYajnik2010,
AgrawalNee2022,ParanjapeSaxena2024}.
\section{Model and radial equations}
\label{sec:model}

We use the standard spherically symmetric monopole ansatz in dimensionless variables,
\begin{equation}
    \phi^a = H(r)\,\hat{x}^a,
    \qquad
    A_i^a = \epsilon_{aij}\frac{x^j}{r^2}\,[1-K(r)] .
\end{equation}
The normalization is chosen so that the metastable broken vacuum is at
\begin{equation}
    H(\infty)=1,
    \qquad
    K(\infty)=0 .
\end{equation}
The radial energy functional is
\begin{align}
\label{eq:energy}
    \frac{E[H,K]}{4\pi}
    =
    \int \dd r
    \Big[
    &\frac12 r^2 H'^2
    +K'^2
    +K^2H^2 
    \\
    &+\frac{(K^2-1)^2}{2r^2} 
    +r^2V_{\rm CW}(H)
    \Big].   
\end{align}

\begin{figure}[t]
    \centering
    \includegraphics[width=\columnwidth]{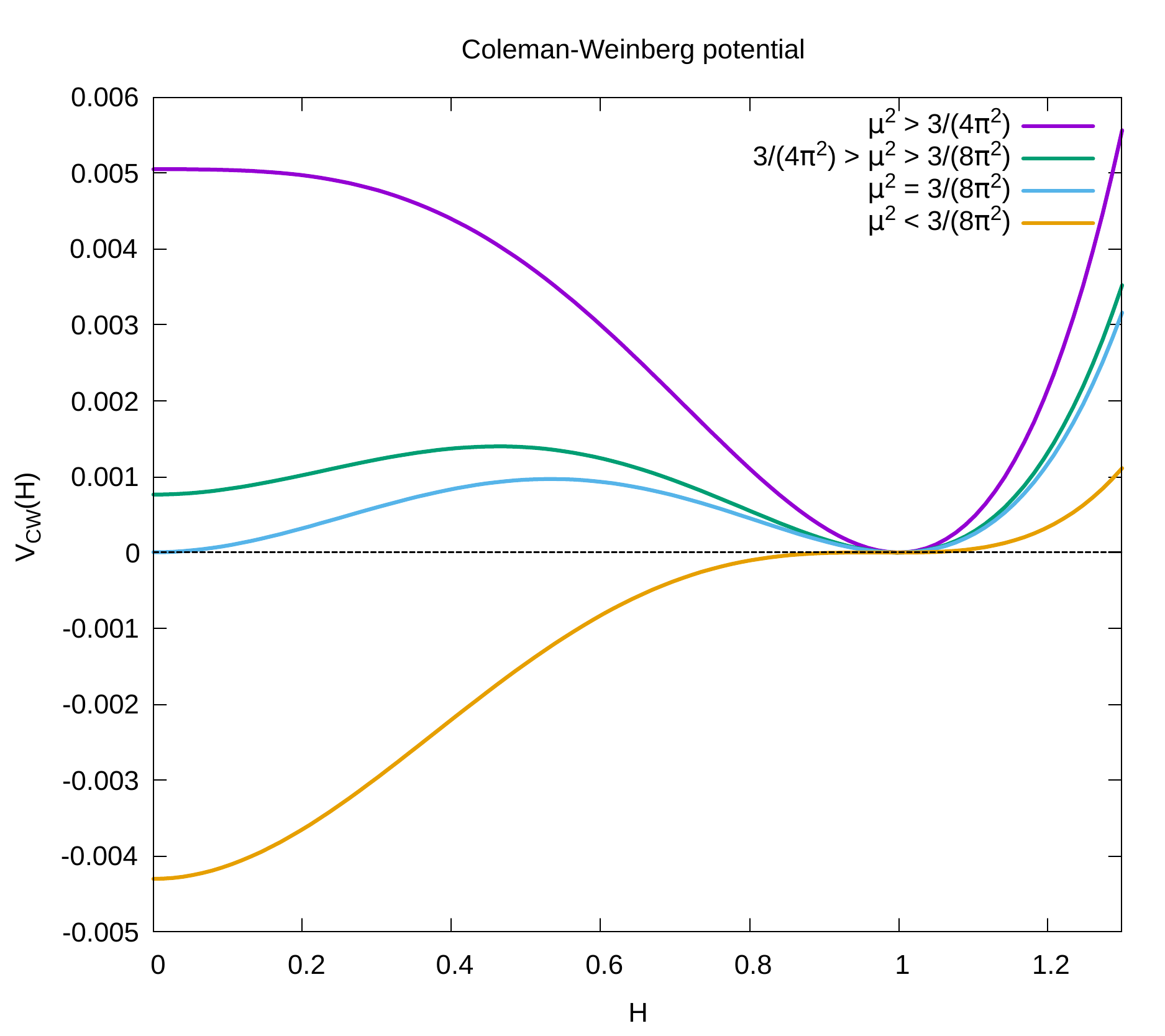}
    \caption{
    Coleman--Weinberg potential used in the radial monopole equations.
    The broken vacuum at \(H=1\) is metastable for
    \(\mu^2<3/(8\pi^2)\), while the lower-energy region lies near \(H=0\).
}
    \label{fig:cw_potential}
\end{figure}

The Coleman--Weinberg potential is taken in the radiatively generated form of ~\cite{ColemanWeinberg1973,Kiselev1990} as

\begin{align}
\label{eq:vcw-kiselev}
V_{\rm CW}(H)
=
\frac{3}{32\pi^2}
\Bigg[
\left(
\frac{4\pi^2}{3}\mu^2-\frac32
\right)&(H^2-1)^2
\nonumber\\
&+H^4\ln H^2
-H^2+1
\Bigg].
\end{align}
This convention gives
\begin{equation}
V_{\rm CW}(1)=0,
\qquad
V_{\rm CW}'(1)=0,
\qquad
V_{\rm CW}''(1)=\mu^2 .
\end{equation}

Thus, \(\mu\) is the rescaled scalar mass around the broken vacuum in the present dimensionless units.  
At the symmetric point,
\begin{equation}
V_{\rm CW}(0)
=
\frac{\mu^2}{8}
-
\frac{3}{64\pi^2}.
\end{equation}
The potential has three qualitatively distinct regimes, as illustrated in Fig.~\ref{fig:cw_potential}.  
For \(\mu^2>3/(4\pi^2)\), the point \(H=0\) is unstable and the broken vacuum at \(H=1\) is the only relevant minimum.  
For
\[
\frac{3}{8\pi^2}<\mu^2<\frac{3}{4\pi^2},
\]
the point \(H=0\) becomes a local minimum, but it lies above the broken vacuum. 
At
\[
\mu^2=\mu_{\rm vac}^2,
\qquad
\mu_{\rm vac}^2=\frac{3}{8\pi^2},
\]
the two vacua are degenerate.  
The metastable regime relevant for this work is therefore
\[
0<\mu^2<\mu_{\rm vac}^2,
\]
where \(H=1\) is a false vacuum and the lower-energy region lies near
\(H=0\).

For the values of \(\mu\) studied below, the ordinary monopole is
supported by the metastable broken vacuum.  
Its core drives the Higgs field away from \(H=1\) toward smaller \(H\), where the lower-energy part of the Coleman--Weinberg potential becomes relevant.  
A finite bubble-like deformation still carries gradient and gauge-field energy, so the monopole--bubble configuration lies above the ordinary monopole branch and appears as a static saddle.  
Denoting the endpoint of the ordinary monopole branch by \(\mu_c\), the static branches considered below lie in the interval
\[
\mu_c < \mu < \mu_{\rm vac}.
\]
The derivative entering the Higgs equation is
\begin{equation}
\label{eq:vcw-prime}
    V_{\rm CW}'(H)
    =
    \frac12\mu^2H(H^2-1)
    +
    \frac{3}{8\pi^2}H
    \left[
    H^2\ln H^2-H^2+1
    \right].
\end{equation}
Varying Eq.~\eqref{eq:energy} gives the coupled radial equations
\begin{equation}
\label{eq:H-eq}
    H''+\frac{2}{r}H'
    -\frac{2K^2H}{r^2}
    =
    V_{\rm CW}'(H),
\end{equation}
\begin{equation}
\label{eq:K-eq}
    K''
    =
    KH^2+\frac{K(K^2-1)}{r^2} .
\end{equation}
The regular origin conditions are
\begin{equation}
    H(0)=0,
    \qquad
    K(0)=1,
\end{equation}
with local expansions
\begin{equation}
\label{eq:origin-expansion}
    H(r)=a r+O(r^3),
    \qquad
    K(r)=1+b r^2+O(r^4).
\end{equation}
At the outer boundary we impose
\begin{equation}
    H(R)=1,
    \qquad
    K(R)=0,
\end{equation}
for the main calculations.  For tail checks we also used the asymptotic Robin condition
\begin{equation}
\label{eq:robin}
    H'(R)+\left(\mu+\frac1R\right)[H(R)-1]=0,
\end{equation}
which follows from
\begin{equation}
    H(r)=1-\frac{A_H}{r}e^{-\mu r}+\cdots .
\end{equation}
The gauge profile decays on the vector scale, $K(r)\sim e^{-r}$, and is numerically negligible at the outer boundary for the boxes used here.

\section{Construction of the saddle}
\label{sec:construction}

The monopole--critical-bubble configuration is a saddle, not a minimum.  A naive gradient flow,
\begin{equation}
    \partial_t U=-\frac{\delta E}{\delta U},
    \qquad U=(H,K), \nonumber
\end{equation}
repels along any Hessian eigenvector with negative eigenvalue.  We therefore construct the solution by a staged Newton method rather than by dissipative relaxation.

First we compute the ordinary Coleman--Weinberg monopole branch by solving Eqs.~\eqref{eq:H-eq}--\eqref{eq:K-eq}.  Second, we compute the scalar $O(3)$ Coleman--Weinberg critical bubble,
\begin{equation}
\label{eq:scalar-bubble}
    H_b''+\frac{2}{r}H_b'=V_{\rm CW}'(H_b),
    ~~
    H_b(r\to\infty)=1 ,
    ~~
    H_b'(0)=0.
\end{equation}
The product
\begin{equation}
\label{eq:product-seed}
    H_{\rm seed}(r)=H_m(r)H_b(r)
\end{equation}
then supplies a useful initial condition: the ordinary monopole profile $H_m$ enforces the regular core, while $H_b$ supplies the exterior bubble deformation.

As an intermediate step, we freeze the gauge field to the ordinary monopole background, $K(r)=K_m(r)$, and solve
\begin{equation}
\label{eq:fixedK}
    H''+\frac{2}{r}H'
    -\frac{2K_m(r)^2H}{r^2}
    =
    V_{\rm CW}'(H).
\end{equation}
This gives a bubble profile in a fixed monopole gauge background.  
Finally, this fixed-background profile is used as the initial condition for the full coupled Newton solve of Eqs.~\eqref{eq:H-eq}--\eqref{eq:K-eq}.

The nonlinear residual is discretized directly from the radial field
equations,
\begin{equation}
    R_H=(r^2H')'-2K^2H-r^2V_{\rm CW}'(H),
\end{equation}
\begin{equation}
    R_K=K''-KH^2-\frac{K(K^2-1)}{r^2}.
\end{equation}
During the Newton solve, we monitor both the maximum residual and a discrete \(L^2\)-type residual over the radial grid.  Since \(R_H\) is written in conservative form, its contribution to these norms is scaled by \(\max(1,r_i^2)\), so that the monitored quantity reflects the local Higgs equation residual rather than the conservative residual multiplied by \(r^2\).  The converged stationary profiles are then used independently in the radial Hessian eigenvalue problem.  
The numerical procedure is only a means of locating the stationary points; the physical diagnostics are the branch energy, \(R_{1/2}\), and the radial Hessian spectrum.

\section{Numerical results and branch structure}
\label{sec:results}

\subsection{Energy and radius branch structure}
\label{subsec:energy-radius}

We characterize the ordinary monopole and monopole--bubble branches by their radial profiles, energies, and a simple radius diagnostic. 
Along each branch we compute
\begin{equation}
    \Efour(\mu)=\frac{E[H,K]}{4\pi},
\end{equation}
and define
\begin{equation}
    H(R_{1/2})=\frac12 .
\end{equation}
For the ordinary monopole, \(R_{1/2}\) measures the monopole core size. For the monopole--bubble saddle, the same quantity tracks the radial position of the bubble-like Higgs transition outside the core.

\begin{figure}[t]
\centering
\includegraphics[width=\columnwidth]{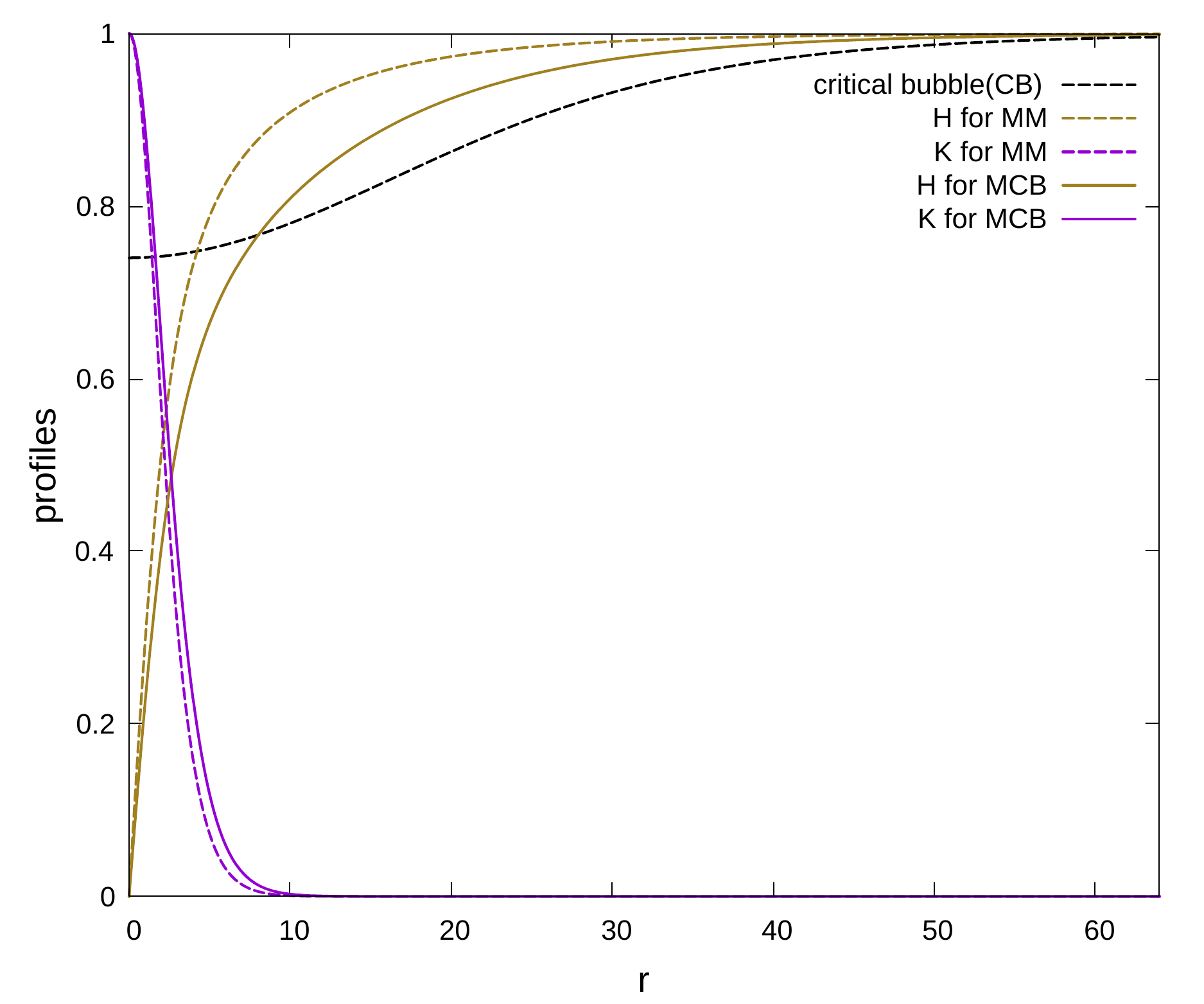}
\caption{
Radial profiles for the metastable monopole (MM), the
monopole--critical-bubble saddle (MCB), and the scalar critical bubble at a representative value \(\mu=0.07\).  
The MCB solution combines a regular monopole core with a bubble-like Higgs transition at larger radius.
}
\label{fig:profiles}
\end{figure}

A representative set of radial profiles is shown in Fig.~\ref{fig:profiles}. 
The full coupled solution keeps the regular monopole core,
\begin{equation}
    H(0)=0,
    \qquad
    K(0)=1,
\end{equation}
while the Higgs field develops a broader transition before returning to the metastable vacuum. 
The gauge field is important near the core, where it regularizes the monopole. 
At larger radii \(K(r)\) has already decayed, and the Higgs equation approaches the scalar critical-bubble equation,
\begin{equation}
    H''+\frac{2}{r}H'=V_{\rm CW}'(H).
\end{equation}
Thus, the saddle combines two ingredients: a regular monopole core and an outer critical-bubble deformation. 

The branches are displayed only up to \(\mu=0.09\) for numerical convenience. 
As \(\mu\) approaches the vacuum-degeneracy value \(\mu_{vac}\), the critical-bubble radius grows rapidly. 
The solution approaches the thin-wall regime, requiring progressively larger radial domains.

\begin{figure}[t]
\centering
\includegraphics[width=\columnwidth]{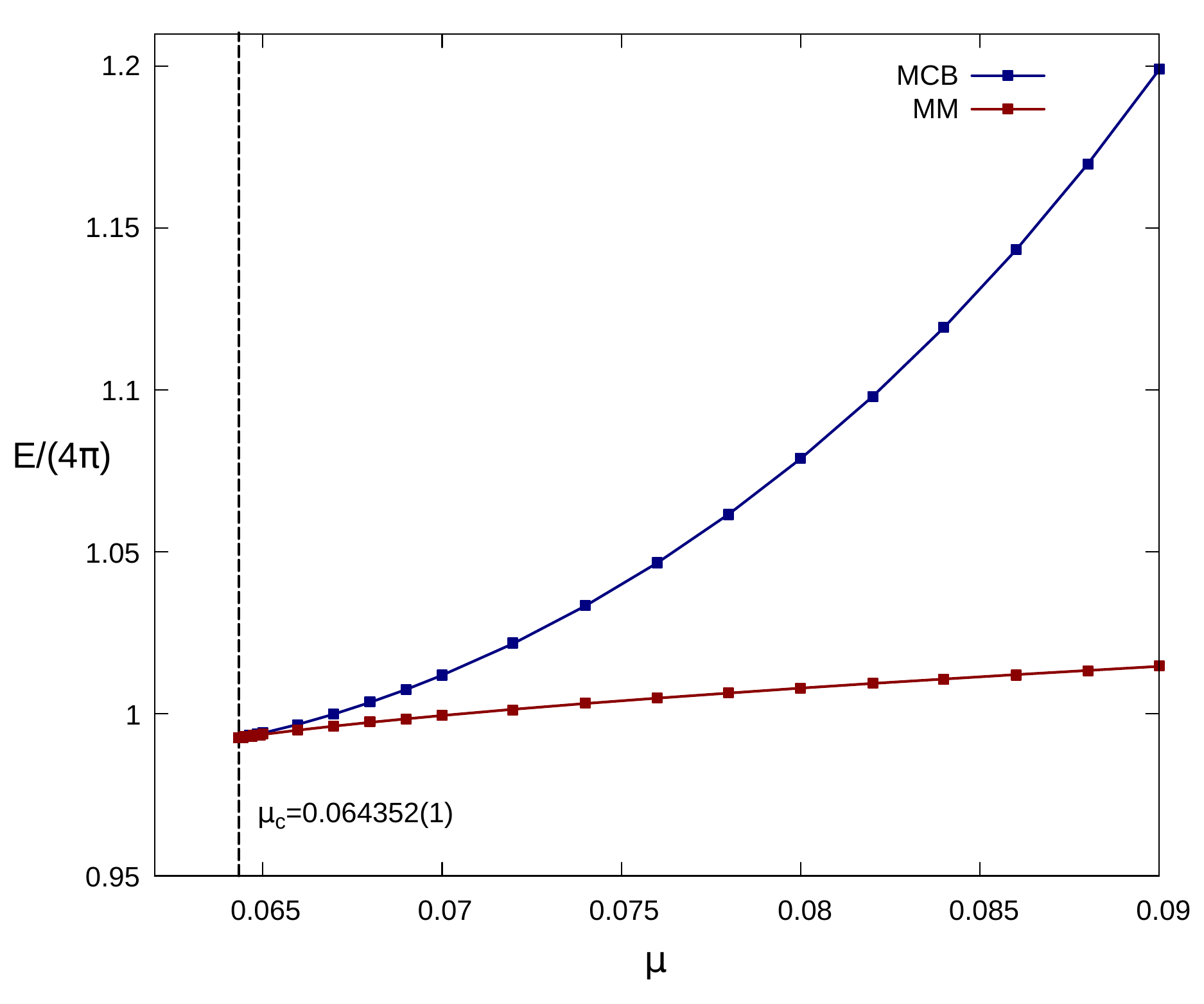}
\caption{
Energy branch structure of the Coleman--Weinberg monopole system.
The critical endpoint is indicated by the vertical line at
\(\mu=\mu_c\).
}
\label{fig:energy-branch}
\end{figure}

Fig.~\ref{fig:energy-branch} shows the energy branch structure. The ordinary monopole is the lower-energy locally stable configuration in the radial sector. 
The monopole--bubble solution lies above it and forms the
associated saddle branch. 
The static barrier along the radial branch can therefore be represented by
\begin{equation}
    \Delta E_{\rm stat}(\mu)
    =
    E_{\rm saddle}(\mu)-E_{\rm mono}(\mu),
\end{equation}
with
\begin{equation}
    \Delta E_{\rm stat}(\mu)>0
\end{equation}
on the part of the branch where both configurations coexist. 
The saddle is not a lower-energy state; rather, it is the critical configuration separating the localized metastable monopole from an expanding-core deformation.

\begin{figure}[t]
\centering
\includegraphics[width=\columnwidth]{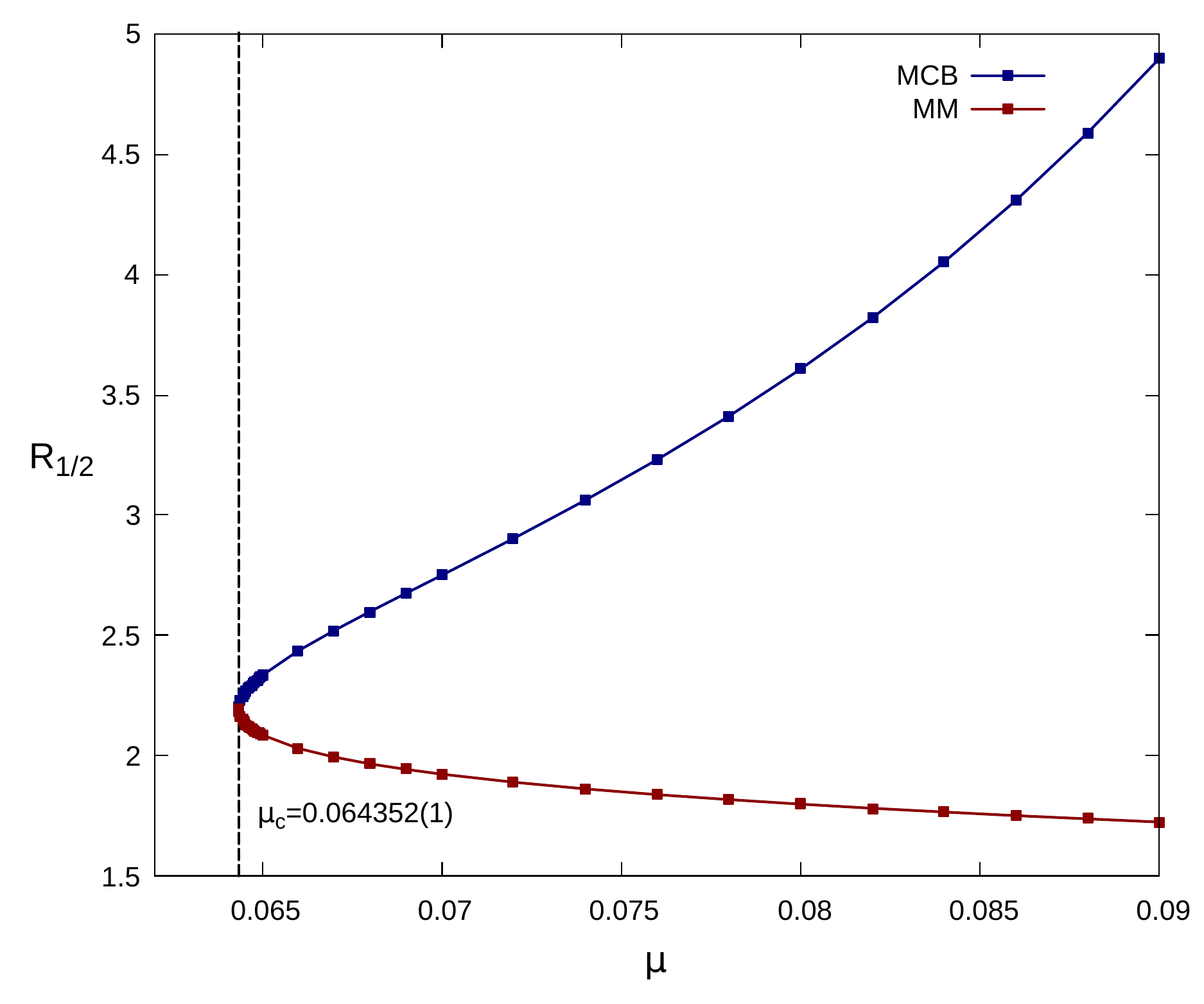}
\caption{
Radius diagnostic \(R_{1/2}\) along the ordinary monopole and
monopole--bubble branches.
}
\label{fig:radius-branch}
\end{figure}

Fig.~\ref{fig:radius-branch} shows the corresponding \(R_{1/2}\)
branch. 
The ordinary monopole branch has the smaller radial scale
associated with the core. 
The saddle branch has a larger \(R_{1/2}\), reflecting the bubble-like Higgs deformation outside the monopole core.
Next we look at the lowest eigenvalue of the radial Hessian and how the Morse character differentiates the regular metastable monopole from the monopole--critical-bubble.

\subsection{Radial Hessian and Morse index}
\label{sec:hessian}

The second variation of Eq.~\eqref{eq:energy} gives the radial Hessian
operator
\begin{equation}
\label{eq:hess-block}
    \mc{H}_{\rm rad}
    =
    \begin{pmatrix}
        \mc{H}_{HH} & \mc{H}_{HK} \\
        \mc{H}_{KH} & \mc{H}_{KK}
    \end{pmatrix},
\end{equation}
where
\begin{equation}
    \mc{H}_{HH}
    =
    -\frac{\dd}{\dd r}
    \left(r^2\frac{\dd}{\dd r}\right)
    +2K^2+r^2V_{\rm CW}''(H),
\end{equation}
\begin{equation}
    \mc{H}_{HK}=\mc{H}_{KH}=4KH,
\end{equation}
and
\begin{equation}
    \mc{H}_{KK}
    =
    -2\frac{\dd^2}{\dd r^2}
    +2H^2+2\frac{3K^2-1}{r^2}.
\end{equation}
The discretized operator is the second derivative matrix of the radial energy functional with the same boundary conditions imposed on the background profiles.
We solve the corresponding shift-invert eigenvalue problem using ARPACK~\cite{Lehoucq1998},
\begin{equation}
\label{eq:hess-eig}
    \mc{H}_{\rm rad}
    \begin{pmatrix}
        \eta_H \\
        \eta_K
    \end{pmatrix}
    =
    \lambda
    \begin{pmatrix}
        \eta_H \\
        \eta_K
    \end{pmatrix}.
\end{equation}
This is the Hessian in the spherically symmetric \((H,K)\) sector.  
It is not the full three-dimensional fluctuation spectrum, and no claim about non-radial modes is made here.

The sign of the lowest eigenvalue gives the radial Morse character of the static solution.  
For the monopole--critical-bubble branch, the lowest radial Hessian mode is negative, showing that this configuration is a
saddle of the radial energy functional.  
As a control, the ordinary Coleman--Weinberg monopole at the same parameter has a positive lowest radial mode and is locally stable within the radial sector.  
The representative eigenvalues and branch diagnostics are summarized in
Table~\ref{tab:representative}.

\begin{table*}[t]
\centering
\caption{
Numerical diagnostics for a representative value \(\mu=0.07\).
}
\label{tab:representative}
\begin{tabular}{lcccccc}
\toprule
configuration & \(\mu\) & \(E/(4\pi)\) & \(R_{1/2}\) & \(\lambda_1\)(lowest) & \(\lambda_2\) & \(\lambda_3\) \\
\midrule
ordinary monopole
& 0.070 & \(0.999628\) & \(1.924453\) & \(0.271547\) & \(1.236251\) & \(1.659331\) \\
monopole--bubble saddle
& 0.070 & \(1.011986\) & \(2.750467\) &  \(-0.434829\) & \(0.727593\) & \(1.339291\) \\
\bottomrule
\end{tabular}
\end{table*}

\begin{figure}[t]
    \centering
    \includegraphics[width=\columnwidth]{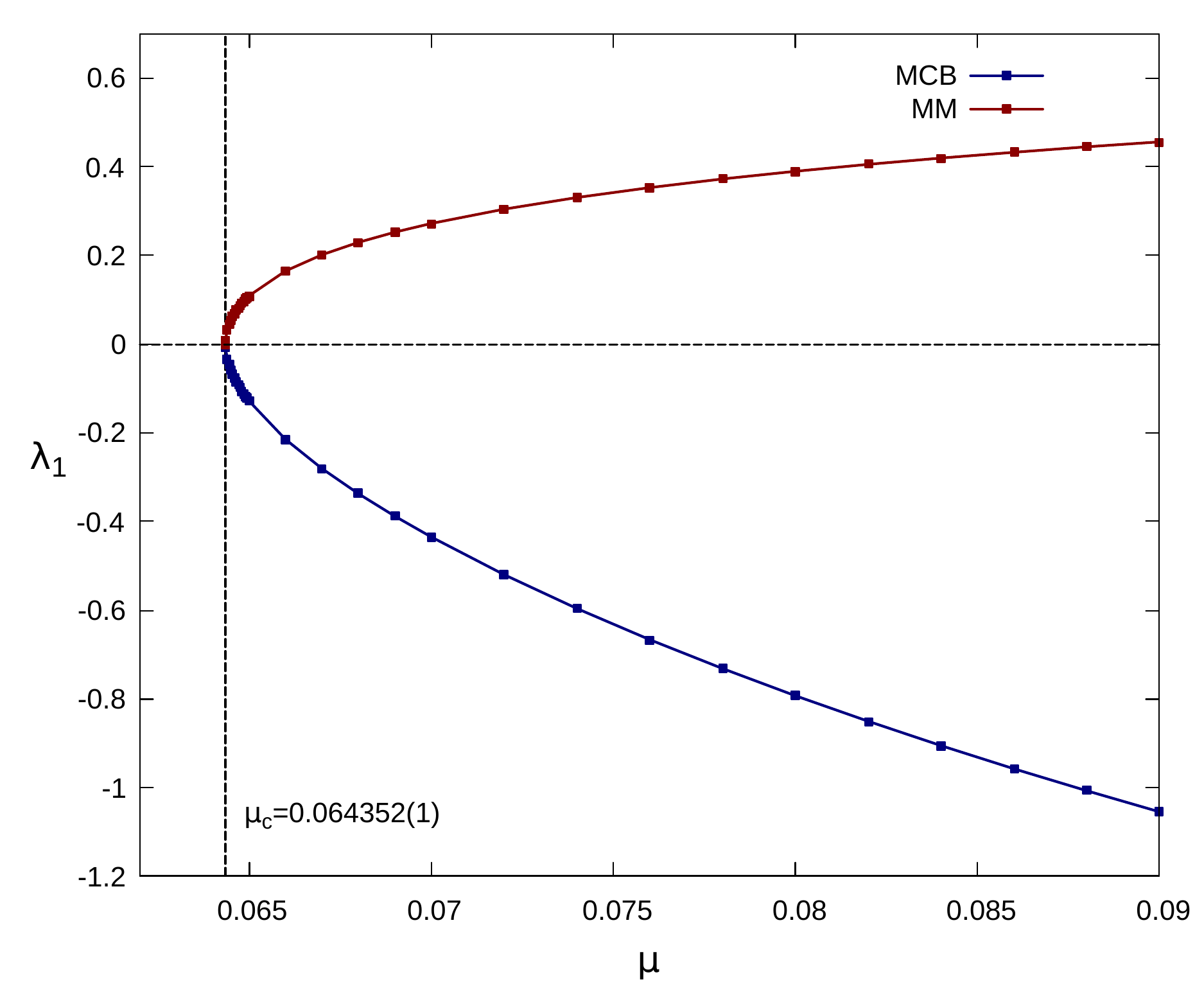}
    \caption{Softening of the lowest radial Hessian eigenvalue near the branch
    endpoint.  The ordinary monopole branch approaches zero from above,
    while the monopole--critical-bubble branch approaches zero from below.}
    \label{fig:softmode}
\end{figure}

The same Hessian diagnostic also identifies the endpoint of the ordinary monopole branch.  
As \(\mu\) is decreased, the lowest radial eigenvalue on the ordinary branch softens and approaches zero. 
This behavior is shown in Fig.~\ref{fig:softmode}.  
The vanishing of \(\lambda_1\) gives a direct spectral criterion for the loss of local radial stability.

\subsection{Branch endpoint from the lowest Hessian mode}
\label{sec:endpoint}

We next follow both the ordinary monopole branch and the monopole--critical-bubble branch toward smaller \(\mu\).  
The endpoint is most clearly identified from the radial Hessian spectrum.  
As the endpoint is approached, the lowest eigenvalue on the ordinary monopole branch approaches zero from above, while the corresponding eigenvalue on the monopole--bubble branch approaches zero from below.  
Thus, the two branches meet with different radial Morse index.

The near-endpoint behavior is shown in Table~\ref{tab:endpoint-hessian}.
The eigenvalues have nearly equal magnitude and opposite sign close to the endpoint, as expected for a stable branch and a saddle branch merging at a fold.

\begin{table}[t]
\centering
\caption{
Lowest radial Hessian eigenvalues near the branch endpoint. 
}
\label{tab:endpoint-hessian}
\footnotesize
\begin{tabular}{ccc}
\toprule
\(\mu\) &
\(\lambda_1^{\rm mono}\) &
\(\lambda_1^{\rm MCB}\) \\
\midrule
\(0.0643530\) &
\(4.32116\times10^{-3}\) &
\(-4.34796\times10^{-3}\) \\
\(0.0643528\) &
\(3.79079\times10^{-3}\) &
\(-3.81145\times10^{-3}\) \\
\(0.0643526\) &
\(3.17248\times10^{-3}\) &
\(-3.18677\times10^{-3}\) \\
\(0.0643524\) &
\(2.39766\times10^{-3}\) &
\(-2.40699\times10^{-3}\) \\
\(0.0643522\) &
\(1.19543\times10^{-3}\) &
\(-1.19675\times10^{-3}\) \\
\(0.06435216\) &
\(7.52878\times10^{-4}\) &
\(-7.53291\times10^{-4}\) \\
\(0.06435214\) &
\(3.60209\times10^{-4}\) &
\(-3.58829\times10^{-4}\) \\
\bottomrule
\end{tabular}
\end{table}

This simultaneous softening identifies the branch endpoint as
\begin{equation}
\label{eq:muc-result}
    \boxed{\mu_c=0.064352(1)}\, .
\end{equation}
Below this value the fixed-\(\mu\) Newton problem becomes singular because the linearized operator develops a zero mode.  The loss of convergence is therefore not used as the definition of the endpoint; rather, the endpoint is defined spectrally by the vanishing of the lowest radial Hessian mode.

Near a fold one expects the leading behavior
\begin{equation}
    \lambda_1^2 \propto \mu-\mu_c .
\end{equation}
Fig.~\ref{fig:softmode} displays this softening. 
The positive branch corresponds to the ordinary monopole, while the negative branch corresponds to the monopole--critical-bubble saddle.
We have also checked the next-lowest radial Hessian eigenvalue along the monopole--critical-bubble branch and find that it remains positive throughout the interval studied within the numerical resolution.  
The monopole--critical-bubble branch therefore has radial Morse index one: it carries exactly one negative radial mode, which softens continuously to zero as \(\mu\to\mu_c^{+}\).

\subsection{Relation to Kiselev's reduced branch parameter}
\label{sec:kiselev}

Kiselev's analysis used a reduced large-distance equation for the
monopole tail~\cite{Kiselev1990}.  
The comparison with that result is controlled by the cubic coefficient of the Coleman--Weinberg potential at the broken vacuum.
Differentiating Eq.~\eqref{eq:vcw-kiselev} gives
\begin{equation}
\label{eq:our-third-deriv}
    V_{\rm CW}'''(1)
    =
    3\mu^2+\frac{3}{2\pi^2}.
\end{equation}
Restoring Kiselev's variables, with \(\phi_0=m/e\), this corresponds to
\begin{equation}
\label{eq:actual-third-deriv}
    V_{\rm eff}'''(m/e)
    =
    \frac{3\mu_K^2 e}{m}
    +
    \frac{3}{2\pi^2}e^3m .
\end{equation}
The reduced asymptotic equation used in ~\cite{Kiselev1990} instead
contains
\begin{equation}
\label{eq:kiselev-third-deriv}
    V_{\rm eff}'''(m/e)
    =
    \frac{3\mu_K^2 e}{m}
    +
    \frac{3}{8\pi^2}e^3m .
\end{equation}
Thus, the loop-induced cubic coefficient entering the reduced tail equation is smaller by a factor of four than the coefficient obtained by directly differentiating the printed potential.  
This coefficient controls the logarithmic correction to the monopole tail and hence the \(C_2\)-branch parameter in Kiselev's calculation.

Kiselev reported the critical value $\mu_c^{\rm K} \simeq 0.0164$, whereas the
full radial calculation gives $\mu_c = 0.064352(1)$. We suspect this
difference arises because Eq.~\eqref{eq:kiselev-third-deriv} is used in the
reduced tail analysis of~\cite{Kiselev1990}, in place of
Eq.~\eqref{eq:actual-third-deriv} obtained by direct differentiation of the
same potential. We note that the ratio \({\mu_c}/{\mu_c^{\rm K}} \simeq 3.92,\)
is numerically close to four, but do not assign any further significance to this proximity.
Kiselev's analysis anticipated the monopole--critical-bubble branch, while here the branch is constructed directly in the full radial
$(H,K)$ system and classified through its Hessian spectrum.

\subsection{Convergence and consistency checks}
\label{sec:checks}

We performed several checks to distinguish the branch structure from discretization or boundary artifacts. 
First, the endpoint of the ordinary monopole branch was monitored under radial grid refinement. 
The endpoint was extracted from the softening of the lowest radial Hessian eigenvalue, rather than from the failure of a fixed-parameter Newton iteration.  
The results are summarized in Table~\ref{tab:grid-muc}.


The corresponding variation is at the level of
\[
    \Delta\mu_c \simeq 6.1\times 10^{-7},
\]
which is small compared with the scale of the branch interval. 
We therefore quote the endpoint conservatively as
\[
    \mu_c = 0.064352(1).
\]
The quoted uncertainty is meant to reflect the observed numerical
variation under refinement, not a statistical error.

\begin{table}[t]
\centering
\caption{
Grid comparison near the soft-mode endpoint for \(R_{\max}=200\).  The lowest radial Hessian
eigenvalue is positive on the metastable monopole branch and negative on
the monopole--critical-bubble branch.
}
\label{tab:grid-muc}
\footnotesize
\begin{tabular}{cccc}
\toprule
\(n_r\) & \(\mu\) &
\(\lambda_1^{\rm mono}\) & \(\lambda_1^{\rm MCB}\) \\
\midrule
\(6001\) & \(0.06435153\) &
\(1.3772\times10^{-4}\) & \(-1.4006\times10^{-4}\) \\
\(24001\) & \(0.06435214\) &
\(3.6021\times10^{-4}\) & \(-3.5883\times10^{-4}\) \\
\bottomrule
\end{tabular}
\end{table}

Second, the residuals of the converged profiles are well below the scale set by the Hessian eigenvalues quoted above.  
This is especially important near \(\mu_c\), where the lowest mode becomes small.  
The branch classification is therefore controlled by the spectrum of stationary solutions, not by residual noise from an incomplete solve.

Third, the same branch picture is obtained from independent diagnostics.
The energy plot shows the monopole--bubble branch lying above the ordinary monopole branch, the \(R_{1/2}\) diagnostic shows the larger radial scale of the bubble-like deformation, and the Hessian spectrum distinguishes the two Morse characters.  
The ordinary monopole branch has a positive lowest radial mode, whereas the monopole--bubble branch carries a negative radial
mode.

These checks give a consistent static picture.  
The ordinary monopole is locally stable in the radial sector until the endpoint is reached, while the monopole--critical-bubble configuration is the corresponding saddle
branch.  
The endpoint is identified by the simultaneous approach of the lowest radial mode to zero on the two branches.

\section{Conclusions}
\label{sec:conclusions}

We have constructed the static Coleman--Weinberg monopole--critical-bubble configuration by solving the full coupled
radial Higgs--gauge boundary-value problem.  
The solution contains a regular monopole core, a bubble-like Higgs deformation at larger radius, and a self-consistently adjusted gauge profile.  
It is therefore a stationary solution of the coupled field equations, not a scalar bubble placed on a fixed monopole background.

The radial Hessian spectrum gives the stability classification.  
Throughout the computed branch interval, the monopole–critical-bubble configuration has one negative radial mode within numerical resolution, showing that it is a saddle of the static energy functional.  
In contrast, the ordinary Coleman--Weinberg monopole has a positive lowest radial mode away from the endpoint and is locally stable within the radial sector.  
The two branches therefore have different Morse character:
the ordinary monopole is the metastable branch, while the monopole--critical-bubble solution is the associated critical configuration.

We also followed the ordinary monopole branch to its endpoint. 
The endpoint is signalled by the lowest radial Hessian eigenvalue going to zero, and our grid-refined estimate gives
\[
    \mu_c = 0.064352(1).
\]
This supports the interpretation that the full radial calculation realizes the branch structure anticipated by the asymptotic analysis.
These results give a direct static picture of how the radially metastable Coleman–Weinberg monopole branch terminates.
The ordinary monopole branch persists down to a soft-mode endpoint, while the associated monopole--critical-bubble branch lies at higher energy and carries the expected negative radial mode. 




\begin{thebibliography}{99}

\bibitem{tHooft1974}
G.~'t Hooft,
Magnetic monopoles in unified gauge theories,
Nucl. Phys. B \textbf{79} (1974) 276.

\bibitem{Polyakov1974}
A.~M. Polyakov,
Particle spectrum in quantum field theory,
JETP Lett. \textbf{20} (1974) 194.

\bibitem{PrasadSommerfield1975}
M.~K. Prasad and C.~M. Sommerfield,
Exact classical solution for the 't Hooft monopole and the Julia--Zee dyon,
Phys. Rev. Lett. \textbf{35} (1975) 760.

\bibitem{Bogomolny1976}
E.~B. Bogomolny,
Stability of classical solutions,
Sov. J. Nucl. Phys. \textbf{24} (1976) 449.

\bibitem{GoddardOlive1978}
P.~Goddard and D.~I. Olive,
Magnetic monopoles in gauge field theories,
Rep. Prog. Phys. \textbf{41} (1978) 1357.

\bibitem{Preskill1984}
J.~Preskill,
Magnetic monopoles,
Ann. Rev. Nucl. Part. Sci. \textbf{34} (1984) 461.

\bibitem{Rajaraman1982}
R.~Rajaraman,
\textit{Solitons and Instantons},
North-Holland, Amsterdam, 1982.

\bibitem{MantonSutcliffe2004}
N.~S. Manton and P.~Sutcliffe,
\textit{Topological Solitons},
Cambridge University Press, Cambridge, 2004.

\bibitem{VilenkinShellard1994}
A.~Vilenkin and E.~P.~S. Shellard,
\textit{Cosmic Strings and Other Topological Defects},
Cambridge University Press, Cambridge, 1994.

\bibitem{Bais2004}
F.~A. Bais,
To be or not to be? Magnetic monopoles in non-Abelian gauge theories,
arXiv:hep-th/0407197.

\bibitem{KobzarevOkunVoloshin1975}
I.~Y. Kobzarev, L.~B. Okun and M.~B. Voloshin,
Bubbles in metastable vacuum,
Sov. J. Nucl. Phys. \textbf{20} (1975) 644.

\bibitem{Coleman1977}
S.~R. Coleman,
The fate of the false vacuum. 1. Semiclassical theory,
Phys. Rev. D \textbf{15} (1977) 2929;
Erratum: Phys. Rev. D \textbf{16} (1977) 1248.

\bibitem{CallanColeman1977}
C.~G. Callan and S.~R. Coleman,
The fate of the false vacuum. 2. First quantum corrections,
Phys. Rev. D \textbf{16} (1977) 1762.

\bibitem{Langer1967}
J.~S. Langer,
Theory of the condensation point,
Ann. Phys. \textbf{41} (1967) 108.

\bibitem{Affleck1981}
I.~Affleck,
Quantum-statistical metastability,
Phys. Rev. Lett. \textbf{46} (1981) 388.

\bibitem{Linde1981}
A.~D. Linde,
Fate of the false vacuum at finite temperature: theory and applications,
Phys. Lett. B \textbf{100} (1981) 37.

\bibitem{Linde1983}
A.~D. Linde,
Decay of the false vacuum at finite temperature,
Nucl. Phys. B \textbf{216} (1983) 421;
Erratum: Nucl. Phys. B \textbf{223} (1983) 544.

\bibitem{Steinhardt1981}
P.~J. Steinhardt,
Monopole and vortex dissociation and decay of the false vacuum,
Nucl. Phys. B \textbf{190} (1981) 583.

\bibitem{PreskillVilenkin1993}
J.~Preskill and A.~Vilenkin,
Decay of metastable topological defects,
Phys. Rev. D \textbf{47} (1993) 2324,
arXiv:hep-ph/9209210.

\bibitem{KumarParanjapeYajnik2010}
B.~Kumar, M.~B. Paranjape and U.~A. Yajnik,
Fate of the false monopoles: induced vacuum decay,
Phys. Rev. D \textbf{82} (2010) 025022,
arXiv:1006.0693.

\bibitem{AgrawalNee2022}
P.~Agrawal and M.~Nee,
The boring monopole,
SciPost Phys. \textbf{13} (2022) 049,
arXiv:2202.11102.

\bibitem{ParanjapeSaxena2024}
M.~B. Paranjape and Y.~Saxena,
Thin-wall monopoles in a false vacuum,
Phys. Rev. D \textbf{110} (2024) 025005,
arXiv:2312.17154.

\bibitem{Kiselev1990}
V.~G. Kiselev,
A monopole in the Coleman--Weinberg model,
Phys. Lett. B \textbf{249} (1990) 269.

\bibitem{EtoHamadaJinnoNittaYamada2022}
M.~Eto, Y.~Hamada, R.~Jinno, M.~Nitta and M.~Yamada,
Abrikosov--Nielsen--Olesen strings from the Coleman--Weinberg potential,
Phys. Rev. D \textbf{106} (2022) 116002,
arXiv:2205.04394.

\bibitem{Kim2024}
E.~Kim,
Large solitons flattened by small quantum corrections,
Phys. Lett. B \textbf{853} (2024) 138681,
arXiv:2405.09262.

\bibitem{ColemanWeinberg1973}
S.~R. Coleman and E.~J. Weinberg,
Radiative corrections as the origin of spontaneous symmetry breaking,
Phys. Rev. D \textbf{7} (1973) 1888.

\bibitem{Lehoucq1998}
R.~B. Lehoucq, D.~C. Sorensen and C.~Yang,
\textit{ARPACK Users' Guide: Solution of Large-Scale Eigenvalue Problems with Implicitly Restarted Arnoldi Methods},
SIAM, Philadelphia, 1998.

\end{thebibliography}
\end{document}